# Molecular Beam Epitaxy grown (Ga,Mn)(As,P) with perpendicular to plane magnetic easy axis

A W Rushforth, M Wang, N R S Farley, R C Campion, K W Edmonds, C R Staddon, C T Foxon, and B L Gallagher

School of Physics and Astronomy, University of Nottingham, University Park, Nottingham, NG72RD, United Kingdom

*We present an experimental investigation of the magnetic, electrical and structural properties of $Ga_{0.94}Mn_{0.06}As_{1-y}P_y$ layers grown by molecular beam epitaxy on GaAs substrates for $y \leq 0.3$. X-ray diffraction measurements reveal that the layers are under tensile strain which gives rise to a magnetic easy axis perpendicular to the plane of the layers. The strength of the magnetic anisotropy and the coercive field increase as the phosphorous concentration is increased. The resistivity of all samples shows metallic behaviour with the resistivity increasing as y increases. These materials will be useful for studies of micromagnetic phenomena requiring metallic ferromagnetic material with perpendicular magnetic anisotropy.*

75.50.Pp, 75.30.Gw, 75.70.Ak

The III-V semiconductor (Ga,Mn)As is one of the most widely studied diluted magnetic semiconductor (DMS) systems exhibiting carrier mediated ferromagnetism and has been utilized in pioneering studies of gate controlled ferromagnetism [1], domain wall resistance [2,3] and current induced domain wall motion [4]. For such studies it is beneficial for the (Ga,Mn)As films to have a perpendicular to plane magnetic easy axis so that magnetisation switching and domain wall motion can be detected using Kerr microscopy [3,4] and the anomalous Hall effect [1,3]. (Ga,Mn)As layers grown directly on GaAs are compressively strained and this leads to in plane magnetic easy axes for Mn concentrations ≥2%, unless the carrier density is very low [5]. (Ga,Mn)As layers under tensile strain, such as when grown on a relaxed InGaAs buffer [6], show a perpendicular to plane magnetic easy axis. However, growth on InGaAs layers can result in a high density of line defects leading to high coercivities and strong pinning of domain walls [7]. Here we demonstrate perpendicular magnetic anisotropy with low switching fields and Tc over 100K, without the requirement of a strain-relaxed buffer layer. This is achieved by incorporation of P at the group V sites during Molecular Beam Epitaxy (MBE) growth of the DMS films, to give large and tuneable tensile strains.

GaP has a smaller lattice constant than GaAs, favouring a stronger exchange interaction when doped with Mn, but has a wider bandgap resulting in stronger localisation of the holes. Experimental investigations on GaP doped with Mn by ion implantation suggest that the Fermi level resides in an impurity band with a relatively low $T_C$=60K [8], while preliminary studies of ion-implanted (Ga,Mn)(As,P) found that $T_C$ was suppressed compared to (Ga,Mn)As [9]. Recently, a theoretical study considered the structural and magnetic properties of (Ga,Mn)(As,P) for a range of P doping levels [10]. Sizeable enhancements of $T_C$ are predicted in this system indicating that the P doping level would be an interesting parameter space for experimental investigations.

Here we present an experimental investigation of the (Ga,Mn)(As,P) system for P levels in the range 10-30%. Three samples were grown by Molecular Beam Epitaxy on [001] GaAs substrates. A 50nm low temperature $GaAs_{1-y}P_y$ buffer layer was grown immediately before a 25nm $Ga_{0.94}Mn_{0.06}As_{1-y}P_y$ layer, each with the same nominal doping y=0.1, 0.2 and 0.3. A control sample was produced with a conventional 25nm $Ga_{0.94}Mn_{0.06}As$ layer grown on a low temperature GaAs buffer. The buffer and Mn containing layers were all grown at the low temperature of 230 $^0$C.

The As flux in our MBE system is provided by a Veeco Mk5 valved cracker set to produce $As_2$. Prior to the main growth the valve setting for As stoichiometry at 580$^0$C is found on a separate test sample, immediately after growth rate calibration using RHEED oscillations.[11]. The stoichiometric point is found by observing the As rich 2x4 to the Ga rich 4x2 surface reconstruction transition[12]. Relative adjustments to this value are then made using the Beam Equivalent Pressure (BEP). There is a higher re-evaporation rate of As at 580$^0$C than at 230 $^0$C and we have determined that, under our conditions, this corresponds to a ~10% excess; thus a 230 $^0$C stoichiometric point can be deduced. The As fluxes for the growth at 230 $^0$C, are chosen with respect to this point, adjusted for the addition of Mn to the group III flux. The phosphorous concentration is established by reducing the As flux from the 230 $^0$C stoichiometric point by the intended phosphorous fraction and then supplying a phosphorous BEP in excess of this fraction (adjusted for the relative phosphorous gauge sensitivity)

The magnetic properties were measured using a Quantum Design MPMS SQUID magnetometer. X-Ray Diffraction (XRD) measurements were obtained using a Philips X'Pert Materials Research diffractometer. Four terminal electrical transport measurements were carried out on Hall bars fabricated using photolithography techniques. Samples were studied before and after annealing in air at $190^0$C for 48 hours. The annealing procedure is an established technique for removing interstitial Mn ions which are the main source of carrier compensation and are detrimental to ferromagnetism in (Ga,Mn)As [13].

For the as-grown samples the Curie temperatures, measured by SQUID magnetometry were 67K, 25K, 13K and 12K for y=0, 0.1, 0.2 and 0.3 respectively. The samples were electrically insulating at low temperatures indicating a high degree of carrier compensation. We shall concentrate mainly on the results for annealed samples. Upon annealing the Curie temperatures recover dramatically as shown in Table 1.

Figure 1(a) shows XRD $\omega$-$2\theta$ scans at the GaAs(004) peak for the annealed films. The broad peak marked with the dashed line corresponds to the thin (Ga,Mn)(As,P) film, and shows a clear shift with increasing P concentration from the low to the high angle side of the sharper peak which corresponds to the GaAs substrate. This shows that the lattice constant decreases with increasing the P concentration and that all the P doped films are under tensile strain. In contrast, the (Ga,Mn)As control sample is under compressive strain. The data were fitted using Philips X'pert Smoothfit software to obtain the values of the perpendicular lattice constant, $a_\perp$, shown in Table I. The P concentrations can be estimated from the XRD data assuming the concentration of Mn and unintentional impurities are the same in the (Ga,Mn)(As,P) films as in the control (Ga,Mn)As film. As shown in Fig 1(b), these are in reasonable agreement with the nominal values, although for the y=0.2 sample the extracted P concentration is somewhat larger (0.27). The fits also indicate that the layer thicknesses are in good agreement with the MBE values, and that the P concentration is slightly higher (by $\Delta y \sim 0.05$) in the Mn-doped films than in the Ga(As,P) buffer layers.

The annealed (Ga,Mn)(As,P) samples show a perpendicular to plane magnetic easy axis with the magnetic anisotropy and coercive field increasing as the P concentration increases (Fig. 2 and Table 1). This is consistent with the increasing tensile strain for increasing P concentration. The increase in the coercive field may also indicate a larger number of defects acting as pinning sites in the material with higher P content. The as-grown samples all have the magnetic easy axis in the plane. This may be a result of the low carrier density in these heavily compensated films [5].

Figure 3 shows resistivity versus temperature curves for the annealed samples. All of the curves show features typical of metallic ferromagnetic (Ga,Mn)As [14]. These are characterised by a steady increase in resistivity as the temperature is decreased from above $T_C$ followed by a sharp drop in resistivity as the temperature is decreased below

$T_C$. The resistivity is finite at low temperatures, but increases as y increases. The as-grown samples all showed insulating behaviour.

GaP has a larger bandgap than GaAs and the Mn acceptor level is deeper in the gap. This may explain the increase in resistivity as y increases if holes are becoming more localised and this may also be responsible for the deterioration of the ferromagnetism. The presence of compensating defects such as As antisites may also lead to such observations. Such defects can form during MBE growth when the As pressure deviates from that required for stoichiometry. The As pressure is a tuneable parameter in the growth of these layers and further refinements may lead to improvements in the magnetic properties of subsequent samples. The close similarity in the magnetic and electrical properties of samples with y=0.2 and y=0.3 indicates that the actual P concentration is similar in these samples, as suggest by XRD measurements. This serves to illustrate the difficulty in successfully incorporating the desired amount of P during the MBE growth.

Our investigations have shown that annealed $Ga_{0.94}Mn_{0.06}As_{1-y}P_y$ layers with y≤0.3 are metallic and are under tensile strain, resulting in a strong perpendicular to plane magnetic anisotropy which increases as y increases. Samples with metallic resistivity and tuneable out of plane magnetic anisotropy and coercive field may find applications in studies of micromagnetic phenomena. For y≤0.1 $T_C$ is comparable to the $Ga_{0.94}Mn_{0.06}As$ control sample. As-grown samples show characteristics of highly compensated material, possibly due to the formation of large numbers of interstitial Mn ions or other compensating defects during growth. To date, the predicted enhancements in $T_C$ [10] in (Ga,Mn)(As,P) have not been observed for our samples.

We acknowledge discussions with Tomas Jungwirth, and funding from EU grant IST-015728 and EPSRC grants GR/S81407/01 and EP/D051487.

Table 1. The Curie temperature $T_C$, perpendicular lattice constant $a_\perp$, magnetic anisotropy field and coercive field for annealed $Ga_{0.94}Mn_{0.06}As_{1-y}P_y$ samples.

Figure 1 (a) X-ray diffraction $\omega$-$2\theta$ scans for the annealed 25nm thick (Ga,Mn)(As,P) films. Grey lines are experimental data, black lines are fits, and the dashed line indicates the 004 peak corresponding to the film. The curves are offset in order of increasing nominal P concentration y, indicated by the numbers shown to the right of the figure. (b) P concentrations extracted from fitting to the curves in (a), assuming that the concentration of Mn and unintentional impurities is the same as for the control sample.

Figure 2 Magnetic hysteresis loops measured at 2K for external magnetic field applied in the plane (closed squares) and perpendicular to plane (open circles) for annealed $Ga_{0.94}Mn_{0.06}As_{1-y}P_y$ layers with (a) y=0.1 and (b) y=0.3.

Figure 3 Resistivity versus temperature for annealed $Ga_{0.94}Mn_{0.06}As_{1-y}P_y$ layers.

| y | $T_C$ (K) | $a_\perp$ (Å) | Anisotropy Field (Oe) | Coercive Field (Oe) |
|---|---|---|---|---|
| 0 | 131 | 5.664 | - | - |
| 0.1 | 110 | 5.639 | 3000 | 30 |
| 0.2 | 80 | 5.608 | 9000 | 170 |
| 0.3 | 77 | 5.600 | 9500 | 170 |

**Table 1**

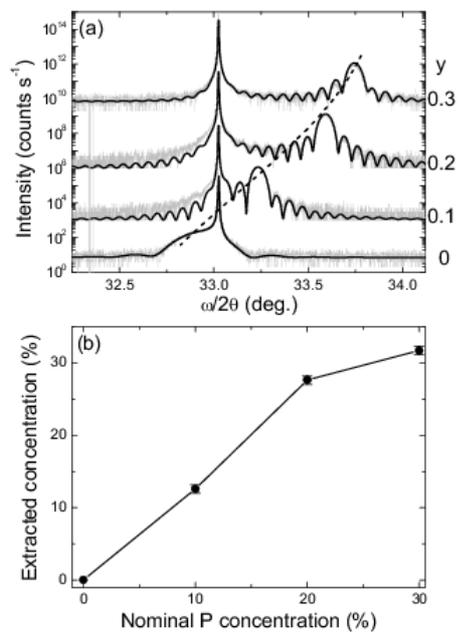

Figure 1

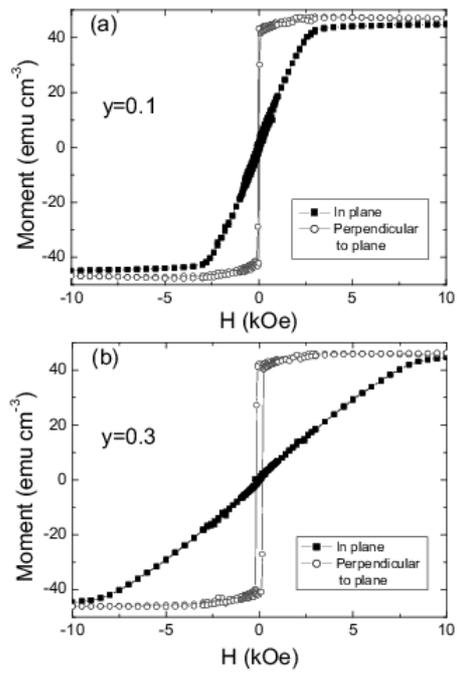

Figure 2

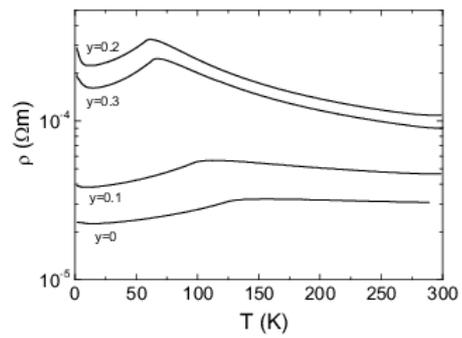

Figure 3